# Tailoring MBE Growth of c-Mn$_3$Sn Directly on MgO (111): From Islands to Film

Longfei He[1,2], Ursula Ludacka[2], Payel Chatterjee[1], Matthias Hartl[1], Dennis Meier[1,2] & Christoph Brüne[1]

[1]Center for Quantum Spintronics, Department of Physics, Norwegian University of Science and Technology, Trondheim, Norway

[2]Department of Materials Science and Engineering, Norwegian University of Science and Technology, Trondheim, Norway

*Corresponding author: longfei.he@ntnu.no, dennis.meier@ntnu.no, christoph.brune@ntnu.no

We present our study of (0001) oriented Mn$_3$Sn (c-Mn$_3$Sn) thin films synthesized directly on an MgO (111) substrate via molecular beam epitaxy. We identify a growth window where Mn$_3$Sn growth can be controlled through slight adjustments of the Mn flux, achieving either µm²-sized high crystalline-quality islands or an almost completely continuous film. High-resolution X-ray diffraction results indicate that both films are highly (0001) oriented. The atomic resolution images show clear film-substrate interfaces displaying an epitaxial relationship. Scanning precession electron diffraction measurements reveal that the island featured sample has highly crystallized Mn$_3$Sn. The sample featuring a high continuity exhibits defects in some areas but retains the dominant Mn$_3$Sn structure. This work demonstrates a potential method for synthesizing high crystalline-quality Mn$_3$Sn films with substantial coverage, facilitating the study of Mn$_3$Sn films without the influence of an additional buffer layer and promoting their application in integrated spintronics.

# Introduction

Kagome antiferromagnetic Mn$_3$Sn attracts significant interest due to its unique physical properties. Its frustrated spin texture results in a non-collinear spin configuration that breaks time-reversal symmetry, inducing magnetic domains[1-3] and potentially forming skyrmions[4]. The topological features in its band structure lead to Weyl nodes that act as monopoles of Berry curvatures[5-9], resulting in novel properties[10-12], such as a large anomalous Hall effect at room temperature[5,13]. Materials in the Kagome family have recently been used in the study of unconventional superconductivity[14-16], thus studying the synthesis of high-quality Kagome thin films is of particular interest. The unconventional band structure make high-quality Mn$_3$Sn thin films an ideal platform for the study of strong electron correlation[17], topology and unconventional superconductivity[18], which ultimately could be a path towards topological quantum computation[14,19]. Additionally, Mn$_3$Sn is antiferromagnetic and thus has negligible net magnetization[20]. These characteristics make Mn$_3$Sn a promising material for research and applications in spintronics as well. To realize this potential, high-quality Mn$_3$Sn thin films are needed. Mn$_3$Sn crystallizes in a hexagonal lattice structure (space group: $P6_3/mmc$) with lattice constants a=5.67 Å and c=4.53 Å (Fig. 1a). Basal planes share an identical lattice structure but are stacked with an offset along the in-plane direction[21]. In each layer, the magnetic Mn atoms form a Kagome lattice structure with each hexagonal center occupied by a Sn atom (Fig. 1b).

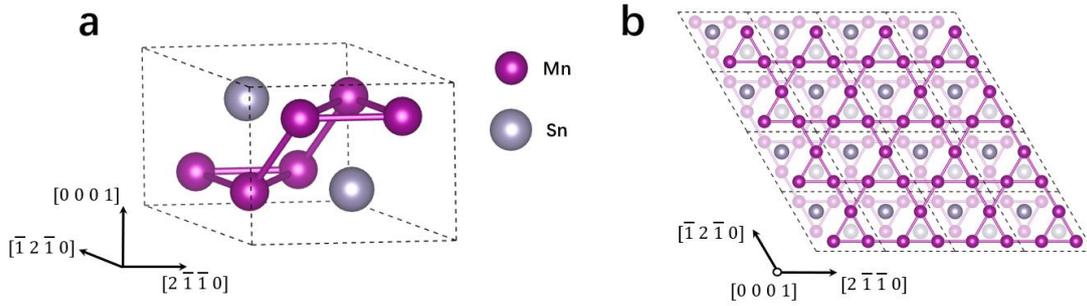

Fig. 1 Crystal structure of Mn$_3$Sn. (a) Mn$_3$Sn unit cell. (b) Mn atoms form a Kagome lattice in each Basal plane.

Synthesizing Mn$_3$Sn in thin film form enables the use of well-developed lithographical processes and techniques for device fabrication in the next step, and it facilitates fine control of film properties such as thickness[22], stoichiometry[23,24] and creating functional interfaces[25]. However, achieving high-quality Mn$_3$Sn thin films remains challenging due to the limited availability of substrates that match Mn$_3$Sn and due to the relatively complex lattice structure of the material. This makes finding suitable epitaxial growth modes challenging especially when searching for layer-by-layer growth conditions that are usually favorable for high-quality, single crystalline thin film growth[26]. Mn$_3$Sn deposition through techniques such as sputtering[7,25,27,28] and pulsed laser deposition (PLD)[29] have been reported, where metal buffer layers, such as Pt or Ru, are commonly employed to minimize lattice mismatch and enhance film quality. However, a metallic buffer layer is often detrimental to transport studies on thin films due to the parallel transport channels it opens up, as well as potential band bending effects at the interface[27,30]. Such effects from the buffer layer complicate the investigations and understanding of Mn$_3$Sn itself. It is therefore important to synthesis high-quality Mn$_3$Sn films without additional metal buffer layers and optimize growth parameters for single crystalline Mn$_3$Sn thin films on insulating substrates, which is one of the main goals of this work. Efforts for depositing high-quality Mn$_3$Sn films on bare Al$_2$O$_3$ substrate by utilizing molecular beam epitaxy (MBE) have been reported[31,32], where the film exhibits relatively large roughness, consist of islands with spherical surfaces, and is limited in size and continuity. The X-ray diffraction result shows multiple out-of-plane lattice orientations of Mn$_3$Sn, reflecting the need for optimized growth conditions.

Here, we report our study of the direct growth of Mn$_3$Sn (0001) thin films on MgO (111) substrates via

MBE. MgO is selected due to its relatively small lattice mismatch with $Mn_3Sn$ (4.7%) compared to other oxide substrates that are available to us. Furthermore, the rather simple surface chemistry of MgO compared with that of, for example, $SrTiO_3$ (111)[33,34] and $Al_2O_3$ (0001)[35], is beneficial, especially for substrate preparation and surface control. For growing high-quality films using MBE, the growth parameters, such as source material beam fluxes and substrate temperature, are of great importance for controlling the growth process. Therefore, we use Knudsen cells to evaporate Mn and Sn source materials separately, which enables control over the individual beam fluxes. Together with the background pressure maintained at $10^{-10}$ Torr during growth, we find a small optimized growth window for the synthesis of $Mn_3Sn$ on MgO. In this growth window, we find the growth of highly (0001) oriented $Mn_3Sn$ with good crystal quality. In addition, a slight decrease of the Mn temperature by 15 °C allows for significant improvements in the film coverage, achieving crystalline $Mn_3Sn$ as an almost completely continuous film. Although a certain amount of degradation in crystallinity is observed in limited areas, the main structure of the $Mn_3Sn$ lattice remains unchanged. SPED analysis indicates that the nearly continuous film is still composed of highly-crystalline $Mn_3Sn$.

# Results

The results presented here are based on growth development and optimization of over 100 samples. To optimize the growth window, we explore a wide range of growth parameters. We focus the discussion on two representative samples that show the two main types of growth results we want to highlight. Sample 1 is synthesized at a substrate temperature of 560 °C, using a Mn evaporation temperature of 755 °C and a Sn evaporation temperature of 980 °C. Sample 2 is synthesized at a substrate temperature of 550 °C with a slightly lower Mn cell temperature during growth of 740 °C, while the Sn temperature remains unchanged. Further details on sample growth can be found in the Experimental section.

**RHEED**

The growth process is in-situ monitored by reflection high-energy electron diffraction (RHEED) for both samples. In Fig. 2a-b, we present a series of RHEED patterns at different growth time. In these images, the electron beam is projecting along the ($1\bar{1}0$) plane with the [111] axis facing upward for MgO, and projecting along the ($11\bar{2}0$) plane with the [0001] axis facing upward for $Mn_{3+x}Sn_{1-x}$. At the start of growth, the RHEED pattern shows the diffraction pattern from the bare MgO (111) substrate. The MgO substrates we use show surface roughness also after substrate preparation. This is also apparent from the RHEED pattern at growth start, which shows signs of transmission diffraction in the form of diffraction dots. We, therefore, index the diffraction dots with their corresponding MgO planes (Fig. 2a).

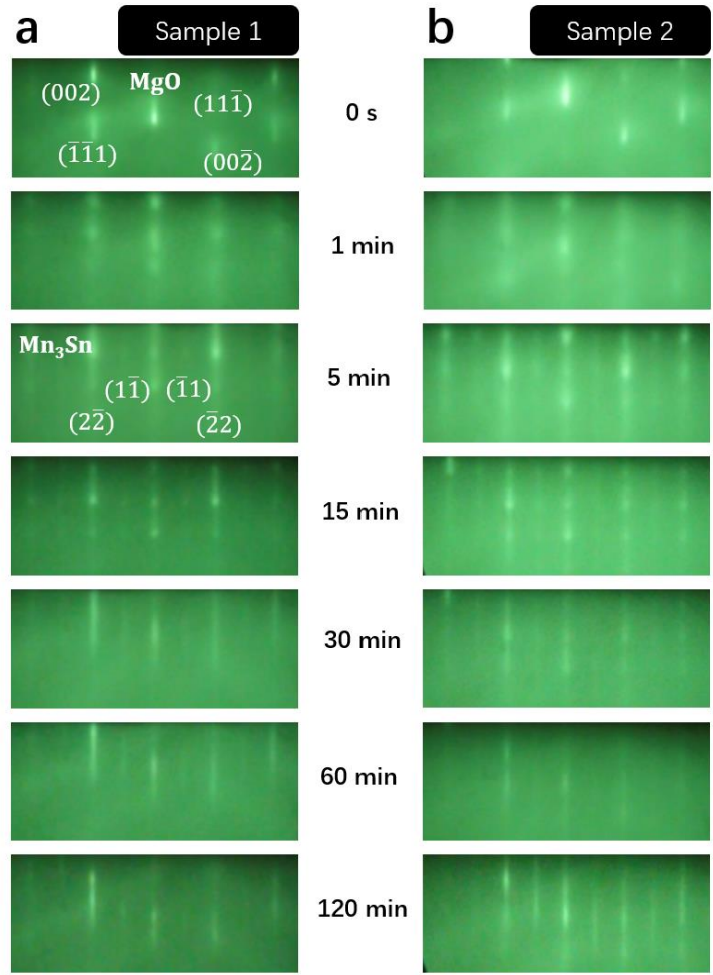

**Fig. 2 RHEED characterizations of film growth.** Rheed evolution during the film growth of Sample 1 (a) and Sample 2 (b). For Sample 1, the RHEED pattern at 0 s is indexed with MgO planes, while the pattern at 5 minutes is indexed based on the reciprocal space vectors of Mn$_3$Sn.

During growth, the RHEED pattern transforms smoothly from the MgO-based pattern to a Mn$_3$Sn-based pattern for both samples. One minute after the growth started, dots from MgO become diffuse and additional dots start to appear, signifying the growth of Mn$_3$Sn. This RHEED pattern indicates that growth at first occurs in a (at least partially) three-dimensional growth mode and not through layer-by-layer growth. This is probably not surprising given the relatively large lattice mismatch and structural/chemical differences between substrate and film material. However, after approximately 5 minutes, the Mn$_3$Sn dots become streaky and start interconnecting along the vertical direction, forming streaks that correspond to the $(2\bar{2})$ and $(\bar{2}2)$ reciprocal space vectors of Mn$_3$Sn; additional streaks that correspond to the $(1\bar{1})$ and $(\bar{1}1)$ reciprocal space vectors also start to appear in between the main streaks, indicating surface reconstruction resulting from crystalline Mn$_3$Sn being synthesized. The RHEED pattern continues to improve for the next 20 to 25 minutes and evolves into a nice streaky RHEED pattern after about 30 minutes for Sample 1. For Sample 2, this transition happens at a later stage around 45 to 60 minutes, which could indicate reduced crystalline-quality or a slower growth rate compared to Sample 1. This is also observed from diffraction and TEM analysis (see below). The streaky RHEED pattern is maintained until the growth is stopped after 120 minutes.

These observations imply that the $Mn_3Sn$ growth is achieved with high crystallinity and predominantly through layer-by-layer growth at the later growth stages. This also signifies the formation of a flat surface of the film/crystallites in the sample. However, we have to note that the RHEED effectively only probes a small surface area, so slow changes in thickness over larger length scales are not visible here. For Sample 2, the RHEED pattern evolves very similar to our observations on Sample 1. However, in the later stage of growth, RHEED on Sample 2 exhibits longer streaks compared to what we see in Sample 1. This potentially indicates growth in larger areas of continuous islands. Additionally, the signal intensity along each streak shows periodic shifts. This could imply small steps or height differences existing on the surfaces in Sample 2, which is investigated in more detail in the next section.

## Morphology

The surface morphology of the as grown samples is investigated using scanning electron microscopy (SEM) and atomic force microscopy (AFM) measurements (Fig. 3). Sample 1 shows islands with smooth surfaces and decent, but incomplete, surface coverage on the substrate as reflected by the SEM image in Fig. 3a. These islands vary in shape and size, separated by gaps in between. Despite their differences, these islands share certain similarities. The edges of the islands are formed by nearly straight lines (Fig. 3a), and the corner of certain islands show relatively regular angles of approximately 60° or 120°, which agrees with the hexagonal lattice structure of $c$-$Mn_3Sn$. In contrast, for Sample 2, a nearly complete surface coverage of the thin film is observed. However, the surface still shows signs of large-scale island formation. The film is significantly improved in terms of feature size and continuity, and the observed gaps in Sample 2 are substantially narrower than in Sample 1.

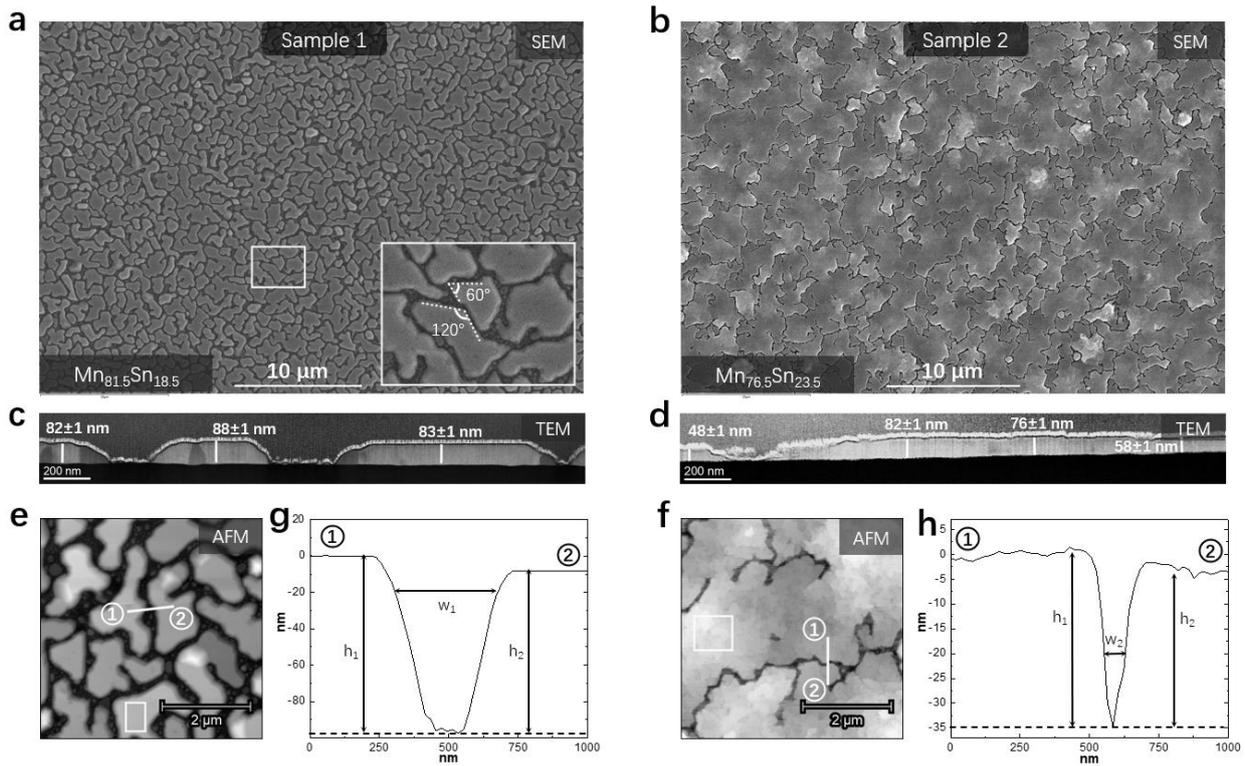

**Fig. 3 Film morphology characterizations and analysis.** (a-b) SEM images of Sample 1 and Sample 2, the inset on bottom left is the stoichiometry measured from EDS. The inset on bottom right of (a) is a zoom in of the area indicated by the square, displaying island features corresponding to the hexagonal lattice structure of $Mn_3Sn$. (c-d) TEM images of Sample 1 and Sample 2, the film thickness is measured at different points as indicated by the lines, with the corresponding height values displayed on top. (e-f) AFM images of Sample 1 and Sample 2, taken with 5x5 µm² scan size. (g) and (h) are line profiles extracted from the lines in (e) and (f), respectively.

Cross section images are taken using transmission electron microscopy (TEM). The islands in Sample 1 (Fig. 3c) display relatively flat surfaces, with slight variation in height both between individual islands and within each island itself. For Sample 2 (Fig. 3d), the height variation is more pronounced across different areas. Within the investigated region, heights range from 48±1 nm to 82±1 nm. Additionally, in the film area of Sample 1, we observe different contrast in different regions, while for Sample 2, the contrast is rather uniform across the entire area. Such darker regions observed in Sample 1 correspond to Mn crystallites, as will be discussed together with the diffraction results later in the paper.

To gain additional information, we perform AFM measurements. Fig. 3e and Fig. 3f are AFM images of Sample 1 and Sample 2, respectively. For Sample 1, the island surfaces show a high degree of flatness, the root mean square of surface roughness (RMS) in the area indicated by the square in Fig. 3e is 1.11± 0.02 nm; while in Sample 2, there are small steps on the surface of individual islands, resulting in a RMS of 3.77±0.20 nm in the area indicated by the square in Fig. 3f. The gaps between islands/areas in each sample are investigated by extracting line profiles in distinct areas (Fig. 3g and Fig. 3h). The line profile presented for Sample 1 indicates a gap depth of about 97±1 nm ($h_1$ in Fig. 3g) and 89±1 nm ($h_2$ in Fig. 3g). These values are comparable to the film thickness measured by TEM (Fig. 3c), suggesting that the islands in the film are fully separated. In contrast, the line profile measured for Sample 2 indicates a much smaller gap size with a depth ⪆ 30 nm ($h_1$ and $h_2$ in Fig. 3h). Although the measured values are below the average film thickness determined based on the cross-sectional TEM data (Fig. 3d), we cannot exclude that the gap extends all the way to the substrate as cavities with a dimension below the physical size of the probe tip are not resolved. In Sample 1, the gap reaches a width of 367±1 nm ($w_1$ in Fig. 3g) at a depth of 20 nm from island surface, while in Sample 2, the gap width develops down to 72±1 nm ($w_2$ in Fig. 3h) at the same depth of 20 nm. Most importantly for this work, the comparison of the AFM line profiles corroborates that Sample 2 has a much higher surface coverage than Sample 1, disrupted only by meandering channels with a width in the sub-hundred nanometers regime.

## HAADF-STEM

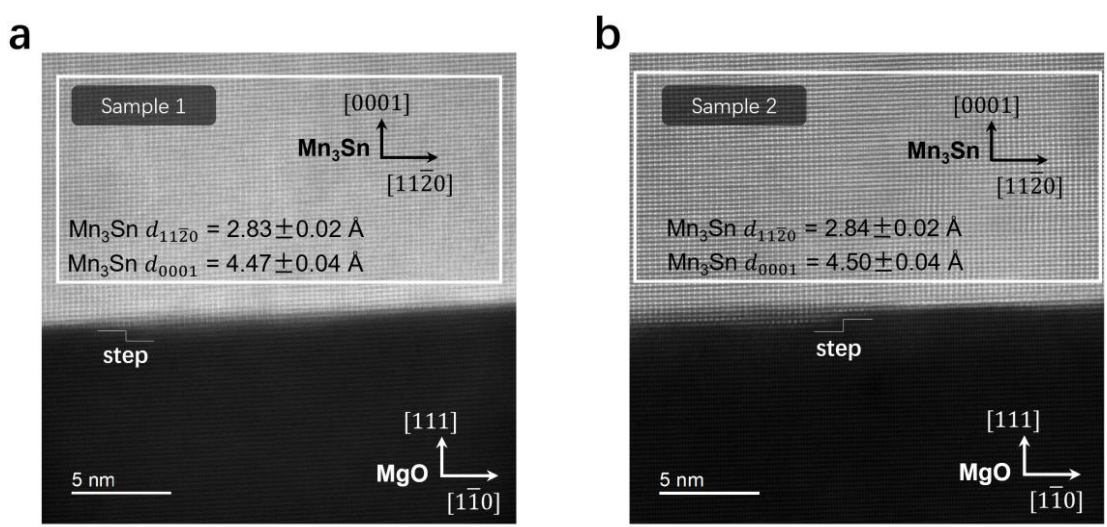

**Fig. 4 Atomic resolution images of the film-substrate interfaces.** HAADF-STEM image of Sample 1 (a) and Sample 2 (b). IFFT is performed on the region indicated by the square to calculate the Mn₃Sn d-spacings as indicated on the bottom left of the square. The film region is indexed with Mn₃Sn lattice orientations determined based on the calculated Mn₃Sn d-spacings.

Cross-sectional high-angle annular dark-field scanning transmission electron microscopy (HAADF-STEM) images (Fig. 4a and Fig. 4b) are taken from both samples along the MgO [11$\bar{2}$] zone axis, with the MgO [111] axis facing upward. For both samples, the bottom dark region corresponds to the MgO substrate, while the top bright region corresponds to the Mn₃Sn film. The film areas show regular positioning of atomic columns along both out-of-plane (vertical) and in-plane (horizontal) directions. The ratio of atomic column spacing along these two directions indicates a lattice structure of Mn₃Sn viewing along the [1$\bar{1}$00] zone axis, with [0001] axis facing upward. Within the region investigated by HAADF-STEM imaging, no large scale of lattice distortion or defects are observed. A clear substrate-film interface is observed and maintains its relative sharpness even across larger surface steps of the MgO substrate (Fig. 4a-b). Inverse fast Fourier transform (IFFT) is performed on the region denoted by the square window in Fig. 4a-b, and the average atomic column spacing is calculated from this. Atomic columns in Sample 1 are separated by 2.83±0.02 Å in the direction parallel to the substrate surface and 4.47±0.04 Å vertically. These values correspond to the expected d-spacing of (11$\bar{2}$0) and (0001) planes in Mn$_3$Sn, respectively. This indicates that the in-plane epitaxy relation between the MgO substrate and the synthesized Mn$_3$Sn films is Mn$_3$Sn [1$\bar{1}$00] // MgO [11$\bar{2}$]. Correspondingly, the atomic column spacing in Sample 2 is $d_{11\bar{2}0}$=2.84±0.02 Å and $d_{0001}$=4.50±0.04 Å.

## Diffraction analysis

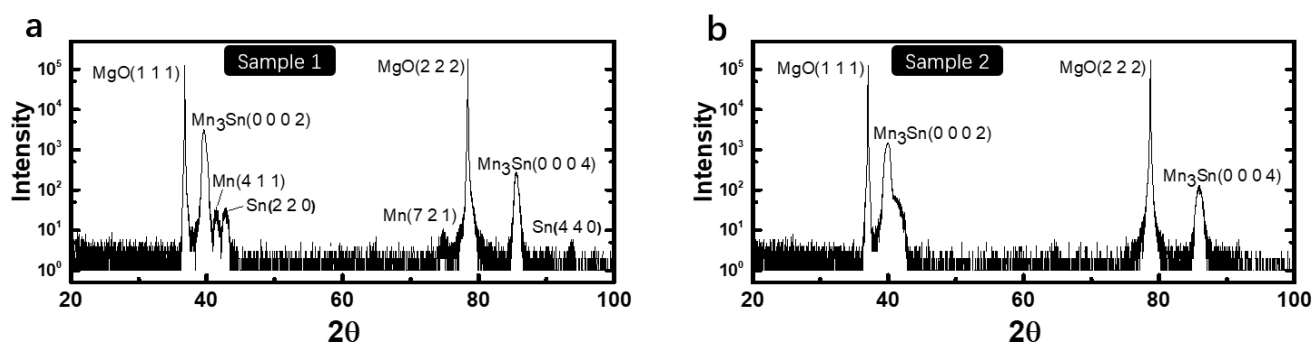

**Fig. 5 HRXRD measurements of synthesized films.** HRXRD pattern of Sample 1 (a) and Sample 2 (b). Both samples show Mn$_3$Sn peaks only from (0002) and (0004) planes. Peaks from Mn and Sn crystallites in Sample 1 are indexed in (a). An uneven broadening is observed on the right shoulder of the Mn$_3$Sn (0002) peak in Sample 2 (b).

To further study the crystallinity, both samples are characterized using diffraction analysis via high-resolution x-ray diffraction (HRXRD) and scanning precession electron diffraction (SPED) measurements. HRXRD of Sample 1 (Fig. 5a) shows Mn$_3$Sn peaks only from (0002) and (0004) planes, which indicates the formation of crystalline c-Mn$_3$Sn in accordance with the HAADF-STEM data. Additional small peaks from Mn and Sn are observed and indexed. These peaks correspond to small amounts of crystalline Mn and Sn inclusions in parts of the film. An example for such an inclusion can also be seen in the TEM data presented in Fig. 3c, where a small section of crystalline Mn can be seen (areas with dark contrast). These Mn inclusions could be induced by the excess Mn being supplied during growth, leading to the formation of crystalline Mn. The Sn crystallites may exist in regions other than the crystalline Mn$_3$Sn and Mn areas, such as in the composites within the island gaps observed in TEM images. The HRXRD of Sample 2 (Fig. 5b) shows peaks from Mn$_3$Sn (0002) and (0004) planes without additional small peaks from Mn or Sn. However, the Mn$_3$Sn peaks become broader, and the Mn$_3$Sn (0002) peak shows signs of an uneven broadening or a secondary peak on its right shoulder.

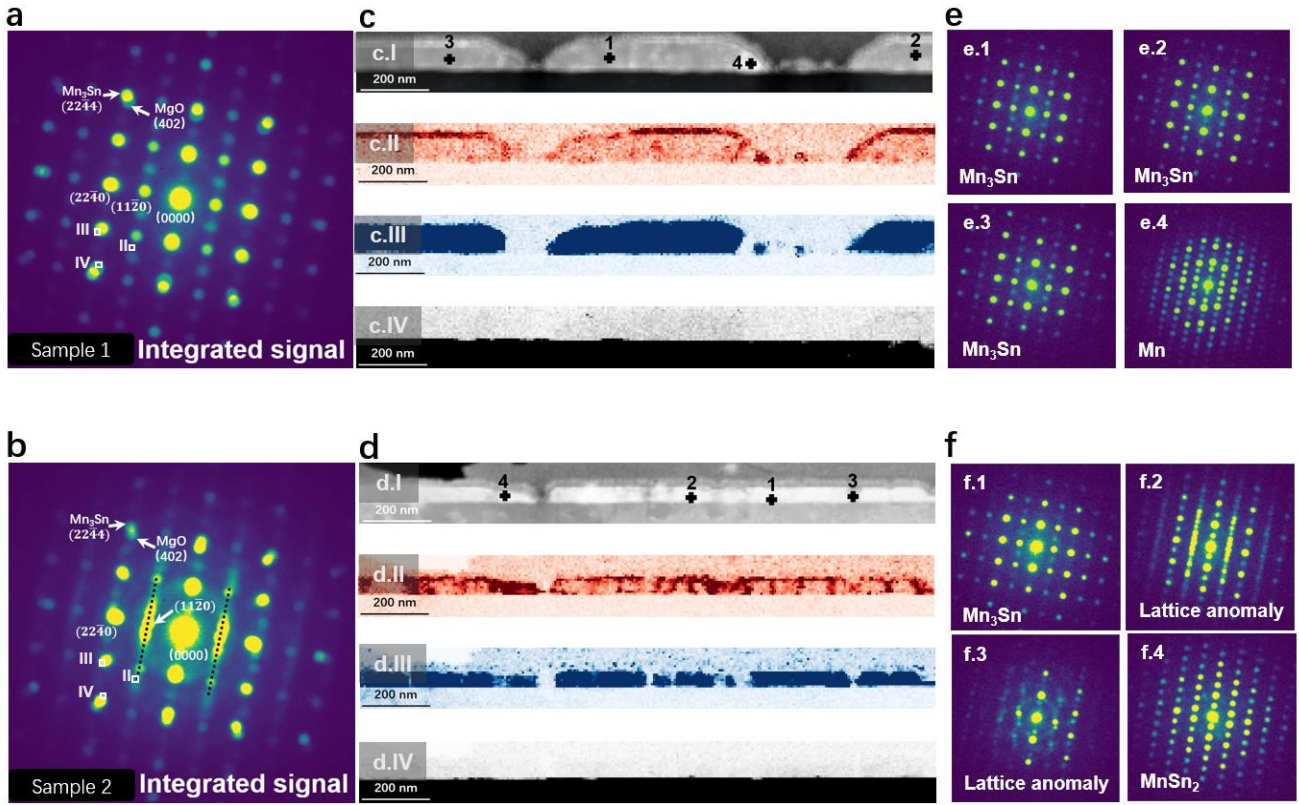

**Fig. 6 Crystallinity analysis based on SPED results.** Integrated diffraction signal of Sample 1 (a) and Sample 2 (b) from the selected area of Sample 1 (c.I) and Sample 2 (d.I), respectively. Lattice anomaly distribution map in Sample 1 (c.II) and Sample 2 (d.II). Mn₃Sn distribution map in Sample 1 (c.III) and Sample 2 (d.III). MgO substrate in Sample 1 (c.IV) and Sample 2 (d.IV). (e) and (f) are diffraction patterns from the selected areas as indicated by the crosses in (c.I) and (d.I), respectively.

We integrate the individual diffraction patterns from SPED into a single diffraction overlay that represents an integrated diffraction pattern for the chosen sample area (as shown in Fig. 6a and Fig. 6b for Sample 1 and Sample 2, respectively. The corresponding selected areas are shown in Fig. 6c.I and Fig. 6d.I). For both samples, we can observe clear diffraction dots whose spacings correspond to the diffraction pattern from $Mn_3Sn$ projecting along the $[1\bar{1}00]$ zone axis. Therefore, dots in Fig. 6a are indexed with their corresponding $Mn_3Sn$ planes. In addition, we observe a weaker diffraction pattern overlaying the $Mn_3Sn$ pattern. This originates from the MgO $[11\bar{2}]$ diffraction and is most apparent for higher order diffraction peaks. For instance, we can clearly observe an MgO (402) dot partially overlapping with the $Mn_3Sn$ ($22\bar{4}4$) dot (marked in Fig. 6a). The integrated signal from Sample 2 (Fig. 6b) shows a similar diffraction structure. The major difference we observe here is the presence of additional dots along the out-of-plane direction (denoted by the dashed line in Fig. 6b) passing through the $Mn_3Sn$ ($11\bar{2}0$) peak.

To investigate the distribution of lattice anomalies in both samples, we examine various locations in the integrated diffraction pattern. We focus on those positions where additional diffraction dots are present in Sample 2 and identify a diffraction feature (denoted as II, marked in Fig. 6a-b), likely arising solely from lattice anomalies. Subsequently, the diffraction signal intensities contributed to this diffraction feature are mapped over the sample (as shown in Fig. 6c.II and Fig. 6d.II for Sample 1 and Sample 2, respectively.). Moreover, we select two additional areas III and IV (marked in Fig. 6a-b) that correspond to crystalline $Mn_3Sn$ and MgO, respectively, then generate their corresponding maps to inspect the film and substrate, comparing the distribution of lattice anomalies systematically.

For both samples, regions of $Mn_3Sn$, lattice anomalies and MgO can be inspected according to the contrast in their corresponding maps. For instance, in Sample 1, we can observe lattice anomalies distributed mainly on the island surfaces as shown in Fig. 6c.II, while crystalline $Mn_3Sn$ contributes to the rest of the dominant regions (Fig. 6c.III). For Sample 2, larger portions of areas are observed with lattice anomalies, including some regions inside the bulk of the film (Fig. 6d.II). This observation agrees with the $Mn_3Sn$ map (Fig. 6d.III), indicating that within the continuous film region, crystalline $Mn_3Sn$ is accompanied by small amounts of lattice anomalies, which are randomly distributed.

Fig. 6e and Fig. 6f are diffraction patterns from selected areas of the sample indicated by the crosses in Fig. 6c.I and Fig. 6d.I, respectively. For both samples, spot 1 represents the typical $Mn_3Sn$ diffraction pattern from undisturbed areas of the film. The d-spacing of $Mn_3Sn$ ($11\bar{2}0$) and $Mn_3Sn$ (0001) planes calculated from the diffraction pattern (Fig. 6e.1 and Fig. 6f.1), is 2.87±0.01 Å and 4.47±0.01 Å for Sample 1, 2.84±0.01 Å and 4.48±0.01 Å for Sample 2. These values agree with the measured values obtained using IFFT on the HAADF-STEM images. For Sample 1, the crystallinity of $Mn_3Sn$ is maintained across different areas, with the diffraction patterns from different spots showing standard diffraction dots from $Mn_3Sn$ (as shown in Fig. 6e.1-3). For Sample 2, the most common lattice anomalies can be presented by the diffraction pattern from spot 2 (Fig. 6f.2), where additional dots appear along the out-of-plane direction passing through the $Mn_3Sn$ ($11\bar{2}0$) dot, while dots belonging to the pure $Mn_3Sn$ lattice remain unchanged. This type of lattice anomaly contributes to the dominant contrast observed in the map of lattice anomalies as shown in Fig. 6c.II and Fig. 6d.II. In addition, we observe another type of lattice anomaly (see Fig. 6f.3). However, this type of diffraction pattern is much less common in our measurements of the film compared to the lattice anomalies represented by spot 2. We see that the $Mn_3Sn$ based diffraction pattern in these areas is heavily distorted, with only a small fraction of the $Mn_3Sn$ signatures remaining.

In addition, we identify small areas of Mn crystallites in Sample 1 (spot 4 in Fig. 6c.I) and $MnSn_2$ crystallites in Sample 2 (spot 4 in Fig. 6d.I). The Mn crystallites observed in Sample 1 likely agree with the HRXRD result, where Mn peaks with different out-of-plane orientations are observed, indicating multi-oriented Mn crystallites incorporated in the film. For Sample 2, although $MnSn_2$ crystallites are observed, no distinct $MnSn_2$ peak appeared in the HRXRD result. This can indicate a very small amount of $MnSn_2$ inclusion. Alternatively, the presence of $MnSn_2$ crystallites may suggest that excess $MnSn_2$, while not in the form of distinct crystallites, is incorporated into the film, leading to lattice imperfections and resulting in the uneven broadening of the Mn₃Sn (0002) peak.

# Discussion

In this study, we have successfully grown highly crystalline, epitaxial thin films of Mn₃Sn directly on MgO substrates. By investigating the dependence on beam flux rates and growth temperatures, we find growth windows that allow for either very flat, high crystalline-quality growth (Sample 1) or nearly full film coverage on the substrate (Sample 2).

Sample 1 demonstrates advancements in several areas compared to previous studies on Mn₃Sn growth using MBE[31,32]. We successfully achieve high-quality thin film growth on MgO (111) substrates without the need for adding Pt or Ru buffer layers as used in previous studies[26,27,29,30]. Growth directly on insulating substrates is generally beneficial for transport studies and optical investigations. Furthermore, the MgO (111) substrates have a smaller lattice mismatch compared to the otherwise commonly used Al₂O₃ substrates and a simpler surface chemistry. This allows for the growth of high crystalline-quality thin films on these substrates as demonstrated here. The Mn₃Sn islands (as shown in Fig. 2a) are achieved with featured shapes that indicate an in-plane hexagonal lattice structure of c-Mn₃Sn, relatively large island sizes and smooth individual island surfaces.

In the HRXRD results (Fig. 5a-b), the Mn₃Sn peaks observed are from (0002) and (0004) planes only, indicating the growth of highly (0001) oriented epitaxial, single crystalline Mn₃Sn. The Mn₃Sn peaks observed in Sample 1 are sharper than those of Sample 2, indicating an agreement with previous findings[24], suggesting that Mn₃Sn films synthesized under Mn-rich conditions exhibit enhanced crystallinity. However, we observe small peaks from Mn and Sn, such as from Mn (411) and Sn (220) planes. These peaks are clearly identifiable in Fig. 5a, indicating that a certain amount of Mn and Sn crystallites are formed in Sample 1. In addition to these excess crystallites, we also find a dominant type of lattice anomaly mostly located on the surface of Sample 1 from the SPED results, indicating that surface reactions like oxidation play an important role. However, there are also weaker contributions discernible from areas inside the bulk of Sample 1. This likely indicates oxidations on the cross-sectional surface, or reduced crystallinity due to growth defects and imperfections in, for example, the stoichiometry of the sample.

For Sample 2, we achieve nearly complete film coverage using adapted growth parameters, where the Mn evaporation temperature was reduced by 15 °C, along with a slight decrease in substrate temperature by 10 °C. The significantly improved film coverage in Sample 2, however, comes at the cost of reduced crystalline-quality. The HRXRD result (Fig. 5b) shows signs of reduced crystallinity in the broadening of the measured Mn₃Sn peaks compared to Sample 1. Additionally, the Mn₃Sn (0002) peak for Sample 2 shows signs of an uneven broadening or a secondary peak on its right shoulder. To further investigate the degradation in crystallinity, we employ SPED (Fig. 6f) to examine the diffraction patterns from the various locations on both samples, through which we identify two types of lattice anomalies that commonly exist. For the dominant type of lattice anomalies (Fig. 6f.2), they are mostly located on the surface for Sample 1 (Fig. 6c.II). For Sample 2, signs for lattice imperfections are found distributed on larger portions of the films as well. The map (Fig. 6d.II) of lattice anomalies suggests that in certain areas, the anomaly extends throughout the entire layer. The second type of lattice anomalies (Fig. 6f.3) is observed only in Sample 2, and such anomalies indicate a degradation in crystallinity. However, its occurrence is much lower than the anomalies represented by spot 2 (Fig. 6f.2). In addition, we observe a small amount of Mn crystallites in Sample 1 and a small amount of $MnSn_2$ crystallites in Sample 2, which can be caused by the larger Mn flux being supplied during the growth of Sample 1. This likely indicates that, with a slight Mn flux adjustment, the growth condition shifted from a Mn crystallite favored (Sample 1) to a $MnSn_2$ crystallite favored (Sample 2) regime.

For both samples, the distribution maps of crystalline $Mn_3Sn$ and lattice anomalies show that highly crystalline $Mn_3Sn$ is making up the majority of the grown film, in agreement to what we find from the HRXRD and EDS analysis. Although Sample 2 exhibits some lattice anomalies within the bulk of the film, they are confined to limited areas, and the regions affected by the most common type of anomalies still retain the dominant $Mn_3Sn$ structure. In return, the film coverage is significantly improved, providing ample area for device or circuit fabrication. This improvement enables a broader exploration of its potential applications.

In conclusion, we present the epitaxial growth of c-$Mn_3Sn$ directly on MgO (111) substrates via MBE. Through our exploration of a wide range of growth parameters, we discover a small growth window where film growth can be efficiently controlled. Within this growth window, adjusting the Mn flux makes it possible to either achieve high crystalline-quality growth of a film of $\mu m^2$-sized $Mn_3Sn$ islands, or $Mn_3Sn$ thin films with high coverage. The high surface coverage growth is achieved by reducing the Mn flux but also results in a reduction in crystallinity. The mechanism within this growth window can be utilized for synthesizing large area $Mn_3Sn$ films with decent crystallinity, or possibly other similar materials with Kagome lattices. Our work enables the direct synthesis of high-quality $Mn_3Sn$ films on insulating substrates. Together with the significantly increased film coverage and continuity, this opens up for transport studies on thin films and the use in more complex layer and device structures (for example, in spin transport and spin diffusion studies).

# Methods

### Synthesis

Both samples were synthesized using MBE. Background pressure in our chamber is typically in the low

$10^{-10}$ Torr range. For the growth, high purity Mn and Sn materials (both 6N) are evaporated from separate Knudsen cells. Before growth, the MgO (111) substrate was pre-annealed at 1000 °C for 1 hour in the growth chamber. The substrate temperature was then lowered to the desired growth temperature and the growth started. The growth parameters are $T_{Mn}$ = 755 °C, $T_{Sn}$ = 980 °C, $T_{Substrate}$ = 560 °C for Sample 1; $T_{Mn}$ = 740 °C, $T_{Sn}$ = 980°C, $T_{Substrate}$ = 550 °C for Sample 2. The growth time for each sample is 2 hours, followed by in-situ post-annealing at the growth temperature for 1 hour. The average thickness of Sample 1 is 90±2 nm according to AFM implying a growth speed of about 0.125 Å/s; for Sample 2, the average thickness is 60±5 nm based on AFM and TEM results, leading to a growth speed of about 0.083 Å/s.

**Measurement**

AFM images are taken by an ICON AFM from Bruker. SEM images were taken from FEI Apreo; film composition was measured by an Oxford EDS that is integrated into the same SEM instrument. The specimens for TEM and SPED measurements were prepared by utilizing a focused ion beam (FIB) Thermo Fisher Scientific G4 UX Dual Beam FIB. A combined system of SEM and a gallium ion beam column were used to prepare the MBE-grown specimens. The two specimens, subsequently extracted from two different samples, were then investigated in a Jeol JEM ARM200F, a double-corrected cold FEG microscope using a high-angle annular dark-field detector (HAADF) at 200 kV. The SPED measurements were conducted on a Jeol 2100F TEM at 200 kV utilizing a Nanomegas P100 scan engine. For data acquisition, we used a Merlin 1S direct electron detector (DED) from Quantum Detectors. The size of the electron beam was set to 2 nm with a convergence angle of 9 mrad. The acquisition parameters included a scan grid of 256 x 256 beam positions with a probe dwell time of 20 ms. HRXRD measurements were conducted using a Cu-K$_\alpha$ ($\lambda$=1.5406 Å) radiation source on a Bruker AXS D8 Discover diffractometer with a half-circle geometry.

# Data availability

The data that support the findings of this study are available from the corresponding authors upon reasonable request.

# Acknowledgements

We thank A. T. J. van Helvoort for discussion. This work was supported by Norwegian University of Science and Technology program, "Enabling Tech" with the project "Novel materials and device design for above room temperature Nano-spintronics", the Research Council of Norway through its Centres of Excellence funding scheme Grant No. 262633 "QuSpin" and its support to the Norwegian Micro- and Nano-Fabrication Facility, NorFab, Project No. 295864.

This work has been submitted for publication.

## Author contributions